\documentclass[12pt]{article}

\usepackage{a4wide}
 
\usepackage{graphicx}
\usepackage{relsize}
\usepackage{tikz}
\usepackage{tikz-3dplot}
\usepackage{amsthm,amsmath,amssymb,bbm}
\usepackage{dsfont} 
\usepackage{enumitem} 
\usepackage{relsize}
\usepackage[utf8]{inputenc}
\usepackage{relsize}
\usepackage[T1]{fontenc} 
\usepackage{lmodern}
\usepackage{caption}
\usepackage{floatrow}
\usepackage{dcolumn}
\usepackage{bm}
\usepackage{color}
\usepackage[normalem]{ulem}
\usepackage{subfig}
\usepackage{mathtools}
\usepackage{scalerel,stackengine}
\usepackage{xcolor}
\usepackage[english]{babel}

\makeatletter

\define@key{x sphericalkeys}{radius}{\def\myradius{#1}}
\define@key{x sphericalkeys}{theta}{\def\mytheta{#1}}
\define@key{x sphericalkeys}{phi}{\def\myphi{#1}}
\tikzdeclarecoordinatesystem{x spherical}{%
    \setkeys{x sphericalkeys}{#1}%
    \pgfpointxyz{\myradius*cos(\mytheta)}{\myradius*sin(\mytheta)*cos(\myphi)}{\myradius*sin(\mytheta)*sin(\myphi)}}

\define@key{y sphericalkeys}{radius}{\def\myradius{#1}}
\define@key{y sphericalkeys}{theta}{\def\mytheta{#1}}
\define@key{y sphericalkeys}{phi}{\def\myphi{#1}}
\tikzdeclarecoordinatesystem{y spherical}{%
    \setkeys{y sphericalkeys}{#1}%
    \pgfpointxyz{\myradius*sin(\mytheta)*cos(\myphi)}{\myradius*cos(\mytheta)}{\myradius*sin(\mytheta)*sin(\myphi)}}

\define@key{z sphericalkeys}{radius}{\def\myradius{#1}}
\define@key{z sphericalkeys}{theta}{\def\mytheta{#1}}
\define@key{z sphericalkeys}{phi}{\def\myphi{#1}}
\tikzdeclarecoordinatesystem{z spherical}{%
    \setkeys{z sphericalkeys}{#1}%
    \pgfmathsetmacro{\Xtest}{sin(\tdplotmaintheta)*cos(\tdplotmainphi-90)*sin(\mytheta)*cos(\myphi)
    +sin(\tdplotmaintheta)*sin(\tdplotmainphi-90)*sin(\mytheta)*sin(\myphi)
    +cos(\tdplotmaintheta)*cos(\mytheta)}
    \pgfmathsetmacro{\ntest}{ifthenelse(\Xtest<0,0,1)}
    \ifnum\ntest=0
    \xdef\MCheatOpa{0.3}
    \else
    \xdef\MCheatOpa{1}
    \fi
    \pgfpointxyz{\myradius*sin(\mytheta)*cos(\myphi)}{\myradius*sin(\mytheta)*sin(\myphi)}{\myradius*cos(\mytheta)}}

\pgfdeclareplothandler{\pgfplothandlercurveto}{}{%
  point macro=\pgf@plot@curveto@handler@initial,
  jump macro=\pgf@plot@smooth@next@moveto,
  end macro=\pgf@plot@curveto@handler@finish
}

\def\pgf@plot@smooth@next@moveto{%
  \pgf@plot@curveto@handler@finish%
  \global\pgf@plot@startedfalse%
  \global\let\pgf@plotstreampoint\pgf@plot@curveto@handler@initial%
}

\def\pgf@plot@curveto@handler@initial#1{%
  \pgf@process{#1}%
  \pgf@xa=\pgf@x%
  \pgf@ya=\pgf@y%
  \pgf@plot@first@action{\pgfqpoint{\pgf@xa}{\pgf@ya}}%
  \xdef\pgf@plot@curveto@first{\noexpand\pgfqpoint{\the\pgf@xa}{\the\pgf@ya}}%
  \global\let\pgf@plot@curveto@first@support=\pgf@plot@curveto@first%
  \global\let\pgf@plotstreampoint=\pgf@plot@curveto@handler@second%
}

\def\pgf@plot@curveto@handler@second#1{%
  \pgf@process{#1}%
  \xdef\pgf@plot@curveto@second{\noexpand\pgfqpoint{\the\pgf@x}{\the\pgf@y}}%
  \global\let\pgf@plotstreampoint=\pgf@plot@curveto@handler@third%
  \global\pgf@plot@startedtrue%
}

\def\pgf@plot@curveto@handler@third#1{%
  \pgf@process{#1}%
  \xdef\pgf@plot@curveto@current{\noexpand\pgfqpoint{\the\pgf@x}{\the\pgf@y}}%
  \pgf@xa=\pgf@x%
  \pgf@ya=\pgf@y%
  \pgf@process{\pgf@plot@curveto@first}
  \advance\pgf@xa by-\pgf@x%
  \advance\pgf@ya by-\pgf@y%
  \pgf@xa=\pgf@plottension\pgf@xa%
  \pgf@ya=\pgf@plottension\pgf@ya%
  \pgf@process{\pgf@plot@curveto@second}%
  \pgf@xb=\pgf@x%
  \pgf@yb=\pgf@y%
  \pgf@xc=\pgf@x%
  \pgf@yc=\pgf@y%
  \advance\pgf@xb by-\pgf@xa%
  \advance\pgf@yb by-\pgf@ya%
  \advance\pgf@xc by\pgf@xa%
  \advance\pgf@yc by\pgf@ya%
  \@ifundefined{MCheatOpa}{}{%
  \pgf@plotstreamspecial{\pgfsetstrokeopacity{\MCheatOpa}}}
  \edef\pgf@marshal{\noexpand\pgfsetstrokeopacity{\noexpand\MCheatOpa}
  \noexpand\pgfpathcurveto{\noexpand\pgf@plot@curveto@first@support}%
    {\noexpand\pgfqpoint{\the\pgf@xb}{\the\pgf@yb}}{\noexpand\pgf@plot@curveto@second}
    \noexpand\pgfusepathqstroke
    \noexpand\pgfpathmoveto{\noexpand\pgf@plot@curveto@second}}%
  {\pgf@marshal}%
  \global\let\pgf@plot@curveto@first=\pgf@plot@curveto@second%
  \global\let\pgf@plot@curveto@second=\pgf@plot@curveto@current%
  \xdef\pgf@plot@curveto@first@support{\noexpand\pgfqpoint{\the\pgf@xc}{\the\pgf@yc}}%
}

\def\pgf@plot@curveto@handler@finish{%
  \ifpgf@plot@started%
    \pgfpathcurveto{\pgf@plot@curveto@first@support}{\pgf@plot@curveto@second}{\pgf@plot@curveto@second}%
  \fi%
}

\def\@fpheader{\relax}

\makeatother

\newcommand{\be}{\begin{equation}}
\newcommand{\ee}{\end{equation}}
\newcommand{\bea}{\begin{eqnarray}}
\newcommand{\eea}{\end{eqnarray}}

\newcommand{\tikzline}[2][fill=black]{\tikz[baseline=-0.5ex] \draw (0,0) circle ; \draw[line width=0.5mm, black] (0,0) -- (0.35,0); \draw (0.35,0) circle ;}%

\usepackage{authblk}

\title{Category-theoretic formulation of relational materialism}

\author[a]{Bekir Bayta\c{s}\thanks{bekirbyts@gmail.com}}
\author[b]{Ozan Ekin Derin\thanks{ozanekinderin@gmail.com}}
\affil[a]{Department of Physics, {\.{I}}zmir Institute of Technology, G{\"{u}}lbah{\c{c}}e, Urla, 35430, {\.{I}}zmir, Turkey}
\affil[b]{Department of Political Science and Public Administration / Graduate School of Social Sciences, Middle East Technical University, \c{C}ankaya,  06800, Ankara, Turkey}
\date{}                     
\begin{document}

\maketitle

\flushbottom


\begin{abstract}

This brief brochure is intended to present a philosophical theory known as relational materialism. We introduce the postulates and principles of the theory, articulating its ontological and epistemological content using the language of category theory. The identification of any existing entity is primarily characterized by its relational, finite, and non-static nature. Furthermore, we provide a categorical construction of particularities within the relational materialist onto-epistemology. Our objective is to address and transform a specific perspective prevalent in scientific communities into a productive network of philosophical commitments.

 \end{abstract}

\section{Basics of relational materialism}
\label{sec:basics-rm}

The relational materialism ($\mathcal{RM}$), as a philosophical standpoint, presupposes the following postulates pertaining to existence and knowledge~\cite{ozbb1}:
\begin{itemize}
\item[i.] \textit{The modes of existence are material}. 
\item[ii.] \textit{An existential mode is material through the presence of all other existential modes.} 
\item[iii.] \textit{The existence of any entity is conditioned upon the possession of existential modes.}
\item[iv.] \textit{Each existential mode has a corresponding mode of knowledge.}
\item[v.] \textit{The modes of knowledge are defined through particular actions performed on the existential modes.}
\end{itemize}

The ontological and epistemological framework of $\mathcal{RM}$ is constructed upon these five postulates.

The general ontology of $\mathcal{RM}$ neither aims to explore a fundamental substance of existence nor to define a single notion encompassing all features of existence; instead, it characterizes entities in terms of their possible existential modes. In this respect, we refer ``being'' of $\mathcal{RM}$ as a category called \textit{\textbf{beable}} ($\mathcal{B}$)\footnote{The term ``beable'' was first used by J. S. Bell to describe the set of elements that may correspond to elements of reality, to things which exist~\cite{bell1}. Our category $\mathcal{B}$ can be seen as the result of a philosophical leverage of this term.}, which can be defined as the \textit{\textbf{likelihood of materiality}} based on the very ontology of $\mathcal{RM}$. Thus, the study of beable allows to extract the possible set of categories and principles of existential modes.

Specifying the criteria of how an entity exists can be followed by conceptualizing the non-existence. The violation of the postulates of $\mathcal{RM}$ regarding existence sets the sufficient and necessary condition for identifying non-existence\footnote{A trivial statement for non-existence is the absence of all modes of the beable, i.e. empty category.}: \textit{Abstracting an existential mode from the beable $\mathcal{B}$}. The action of abstraction is a logical operation that severs the coexistency of the existential mode from other modes. The abstraction of existential modes results in the classes of the category \textit{\textbf{non-beable}} ($\mathcal{NB}$), which naturally belongs to the study of \textit{\textbf{immateriality}}.

The general epistemology of $\mathcal{RM}$ focuses on the general existential conditions of knowledge. In the context of the fourth postulate of $\mathcal{RM}$, the ontological categories of $\mathcal{RM}$ are integrated within the categories of $\mathcal{RM}$-epistemology, where every mode of knowledge has a corresponding relationship with the integrated mode of existence. Consequently, the possible set of modes of knowledge is constrained by $\mathcal{B}$. We can introduce a corresponding category of $\mathcal{B}$ in $\mathcal{RM}$-epistemology, termed \textit{\textbf{knowable}} ($\mathcal{K}$). As $\mathcal{B}$ refers to the likelihood of materiality in $\mathcal{RM}$-ontology, the category $\mathcal{K}$ can be associated with the \textit{\textbf{likelihood of knowability}} in $\mathcal{RM}$-epistemology. Similar to the categories of $\mathcal{NB}$, the abstraction of the modes of knowledge provides the categories of \textit{\textbf{non-knowable}} ($\mathcal{NK}$).

The general ontology of $\mathcal{RM}$ and the general epistemology of $\mathcal{RM}$ provide the \textit{\textbf{universal modes}} of existence and knowledge, respectively (with the latter being a specific form of the former). At the level of $\mathcal{RM}$-ontology, we have assumed a form of \textit{\textbf{multiplicity}} for the beable $\mathcal{B}$ and for the knowable $\mathcal{K}$. However, neither $\mathcal{RM}$-ontology nor $\mathcal{RM}$-epistemology can address the set of possible \textit{\textbf{particular}} modes of existence and their corresponding modes of knowledge. Particular modes of existence are articulated with respect to the modes of general ontology, thereby forming ontological subcategories. Their existential states also have a corresponding set of modes of knowledge. Henceforth, we refer to the categorical universe of the particular modes as the \textit{\textbf{onto-epistemological particularities}} of $\mathcal{RM}$, labelled by $\mathcal{P}$.

The inquiry into the concrete identification of each element of the modes of existence and knowledge is an \textit{\textbf{a posteriori}} question. Their concrete ascertainability is achieved through our practices of comprehending our environment. The practices to which we refer are, by nature, the \textit{\textbf{scientific}} ones, which, thus far, represent the most credible method for explaining the nature of phenomena. One or multiple concrete modes may manifest in any scientific practice. Consequently, we undertake an abstraction to elevate these concrete modes to a (sub)category within $\mathcal{RM}$.

In what follows, we present a concise formulation of essential features in $\mathcal{RM}$, employing insights from category theory~\cite{saunders, steve} to elucidate the interrelations and transitions of concepts within this philosophical framework. We utilize categories, their classes, morphisms, and functors to articulate the propositions and connections in $\mathcal{RM}$ in terms of the language of category theory.

\section{Relational materialist onto-epistemology}
\label{sec:ontoepis-rm}

$\mathcal{RM}$-ontology and $\mathcal{RM}$-epistemology set forth the principles of beable $\mathcal{B}$ and knowable $\mathcal{K}$, respectively. The fourth postulate of $\mathcal{RM}$ states that there is a correspondence between the modes of existence and the modes of knowledge. We prefer to call these modes, which form the general and particular categories of existence and knowledge, as \textit{onto-epistemological} modes.

There are three universal onto-epistemological modes in $\mathcal{RM}$: 
\begin{equation}
1. \,\, \mathrm{\textit{Relationality}} \equiv \mathcal{R}\,, \quad 2. \,\, \mathrm{\textit{Finitude}} \equiv \mathcal{F}\,, \quad  3. \,\,\mathrm{\textit{Non-staticity}} \equiv \mathcal{NS}\,.
\end{equation}

Each of these universal modes plays a particular role of an establisher principle for both $\mathcal{B}$ and its corresponding category $\mathcal{K}$. In this respect, all universal onto-epistemological modes are labelled by $\mathcal{B}$ or $\mathcal{K}$ depending on what type of principle they specify and their presence is due to being constituents of $\mathcal{B}$ and $\mathcal{K}$.

Relationality, finiteness, and non-staticity characterize all entities and their existential features within this framework. Each of these modes cannot adequately qualify the beable $\mathcal{B}$ in a coherent manner unless they are \textit{\textbf{coexistent}} within $\mathcal{B}$. Specifically, entities exist through relationality; without relationality, they cannot be finite. Similarly, non-static entities cannot be studied in the absence of relationality, as their dynamism is possible through it. Therefore, the absence of any one of ($\mathcal{R}, \mathcal{F}, \mathcal{NS}$) for any potential entity is sufficient to negate the existence of that entity. In essence, absoluteness ($\mathcal{A}$), infinitude ($\mathcal{IF})$, and staticity ($\mathcal{S}$) merely define the forms of non-beable $\mathcal{NB}$, which are formal negations of $\mathcal{R}$, $\mathcal{F}$ and $\mathcal{NS}$, respectively and they do not necessarily belong to the domain of the onto-epistemology of $\mathcal{RM}$.

\subsection{General ontology of $\mathcal{RM}$}

Consider the beable $\mathcal{B}$ as a category with the following collection of objects, morphisms and related features:
\begin{itemize}
\item[1.] The objects of $\mathcal{B}$ are given by
\be
\mathrm{ob}(\mathcal{B}): \,\, \mathcal{R}_{\mathcal{B}}, \mathcal{F}_{\mathcal{B}}, (\mathcal{NS})_{\mathcal{B}} \,, 
\ee
where each onto-epistemological mode ($\mathcal{R}, \mathcal{F}, \mathcal{NS}$) in $\mathrm{ob}(\mathcal{B})$ is labelled by the beable $\mathcal{B}$ and each can be represented as an ontologically irreducible class of $\mathcal{B}$. 
\item[2.] $\mathcal{B}$ is a \textit{discrete} category, in which the morphisms of $\mathcal{B}$ are only the identify maps,
\be
m_{\mathcal{B}}(X,X) = \mathrm{id}_{X} \,, \,\,\,\, \forall X \in \mathrm{ob}(\mathcal{B}) \quad \mathrm{and} \quad m_{\mathcal{B}}(X,Y) = \varnothing \,, \,\,\,\,  \forall \, X \neq Y \,.
\ee
This is to emphasize that there is no morphism that maps an object in $\mathrm{ob}(\mathcal{B})$ to any other object in $\mathrm{ob}(\mathcal{B})$. 
\item[3.] $\mathcal{B}$ is a self-dual category $\mathcal{B}^* \cong \mathcal{B}$ in the sense that:
\be
\mathrm{ob}(\mathcal{B}^*) = \mathrm{ob}(\mathrm{\mathcal{B}}) \quad \mathrm{and} \quad m_{\mathcal{B}^*}(X,Y) = m_{\mathcal{B}}(X,Y) \,,  \quad  \forall X, Y \in \mathrm{ob}(\mathcal{B}) \,,
\ee
as the morphisms $\mathrm{ob}(\mathrm{\mathcal{B}})$ of $\mathcal{B}$ are composed of identity maps $m_{\mathcal{B}}(X,Y) =  \mathrm{id}_{X}$ when $X =Y$, otherwise it is empty.
\end{itemize}


The initial object of categories $\mathrm{cat}(\mathcal{NB})$ of non-beable $\mathcal{NB}$ is the empty category $\varnothing_{\mathcal{NB}}$, which represents the absence of all universal modes and no morphisms. The other classes of non-beable $\mathcal{NB}$ are obtained by the action of a functorial lift called \textit{abstraction} ($A$) on the objects of $\mathcal{B}$ that ``isolates'' each object in $\mathrm{ob}(\mathrm{\mathcal{B}})$ from $\mathcal{B}$, 
\be
A: \mathrm{ob}(\mathcal{B}) \mapsto \mathrm{ob}(\mathcal{X}_{\mathcal{NB}}) \,, \quad \mathcal{X}_{\mathcal{NB}} \in \mathrm{cat}(\mathcal{NB}) \,,
\ee
and maps between morphisms in $\mathcal{B}$ and $\mathcal{X}_\mathcal{NB}$:
\be
A: m_{\mathcal{B}}(X,X) \mapsto m_{\mathcal{X}_\mathcal{NB}}(A(X), A(X)) \,, \quad \forall X \in  \mathrm{ob}(\mathcal{B})\,,
\ee
where the categories $\mathcal{X}_{\mathcal{NB}} \in \mathrm{cat}(\mathcal{NB})$ are discrete monoids:
\be
\mathrm{cat}(\mathcal{NB}): \,\, \mathcal{A}, \, \mathcal{IF}, \, \mathcal{S}.
\ee

The functor $A$, as a logical operation, induces a \textit{mono-morphism}: $\mathrm{ob}(\mathcal{B}) \mapsto \mathrm{ob}(\mathcal{X}_{\mathcal{NB}})$ and $m_{\mathcal{B}}(X,X) \mapsto m_{\mathcal{X}_{\mathcal{NB}}}(A(X), A(X))$, where each object in $\mathrm{ob}(\mathcal{B})$ is mapped to its negation in $\mathrm{ob}(\mathcal{X}_{\mathcal{NB}})$ and each identity map $\mathrm{id}_X$ in $m_{\mathcal{B}}(X,X)$ is mapped into the corresponding identity map $\mathrm{id}_{A(X)}$ in $m_{\mathcal{X}_{\mathcal{NB}}}(A(X),A(X))$. Moreover, the categories of $\mathrm{cat}(\mathcal{NB})$ are essentially isomorphically equivalent:
\be
\exists \, F_{\mathcal{X}_{\mathcal{NB}},\mathcal{Y}_{\mathcal{NB}}}, G_{\mathcal{X}_{\mathcal{NB}},\mathcal{Y}_{\mathcal{NB}}} \quad \mathrm{s.t.} \quad \mathcal{X}_{\mathcal{NB}} \,\, \xrightleftharpoons[G_{\mathcal{X},\mathcal{Y}}]{F_{\mathcal{X},\mathcal{Y}}} \,\, \mathcal{Y}_{\mathcal{NB}}\,, \quad \forall \, \mathcal{X}_{\mathcal{NB}},\mathcal{Y}_{\mathcal{NB}} \in  \mathrm{cat}(\mathcal{NB})\,, 
\ee
such that the functors $F_{\mathcal{X},\mathcal{Y}}, G_{\mathcal{X},\mathcal{Y}}$ are related: $G_{\mathcal{X},\mathcal{Y}} \, \circ \, F_{\mathcal{X},\mathcal{Y}} = \mathrm{id}_{\mathcal{X}_{\mathcal{NB}}}$ and $F_{\mathcal{X},\mathcal{Y}} \, \circ \,G_{\mathcal{X},\mathcal{Y}} = \mathrm{id}_{\mathcal{Y}_{\mathcal{NB}}}$. This shows that the objects of $\mathrm{cat}(\mathcal{NB})$ are equivalent representations of the non-existence: $\mathcal{A} \cong \mathcal{IF} \cong \mathcal{S}$. Therefore, the resulting action of the abstraction functor is that the objects of $\mathcal{B}$ cannot be present as an existential mode within the ontological framework of $\mathcal{RM}$.

\subsection{General epistemology of $\mathcal{RM}$}

There exists an injection (faithful) functor $I_{\mathcal{B} \mathcal{K}}$ that maps the objects of $\mathcal{B}$ into the objects of the category $\mathcal{K}$, in the sense that $I_{BK}$ maps $\mathcal{B}$ to its equivalent full subcategory $I_{\mathcal{K}} \, \mathcal{B} \cong \mathcal{K}_{\mathcal{B}}$ in $\mathcal{K}$:
\be
\exists \, I_{\mathcal{B} \mathcal{K}}, \,\,\,  I_{\mathcal{B} \mathcal{K}}: \, \mathrm{ob}(\mathcal{B}) \mapsto \mathrm{ob}(\mathcal{K}_{\mathcal{B}}) \,,
\ee
where the objects $\mathrm{ob}(\mathcal{K}_{\mathcal{B}})$ and the morphisms $m_{\mathcal{K}_\mathcal{B}}(I_{\mathcal{B} \mathcal{K}}(X), I_{\mathcal{B} \mathcal{K}}(X))$ are given by
\be
\mathrm{ob}(\mathcal{K}_{\mathcal{B}}):  \,\, \mathcal{R}_{\mathcal{K}_{\mathcal{B}}}, \mathcal{F}_{\mathcal{K}_{\mathcal{B}}},  \mathcal{(NS)}_{\mathcal{K}_{\mathcal{B}}} \quad \mathrm{and} \quad m_{\mathcal{K}_\mathcal{B}}(I_{\mathcal{B} \mathcal{K}}(X), I_{\mathcal{B} \mathcal{K}}(X)) = \mathrm{id}_{I_{\mathcal{B} \mathcal{K}}(X)}\,,
\ee
which implies that the beable $\mathcal{B}$ is embedded as a subcategory into the knowable $\mathcal{K}$.

The subcategory $\mathcal{K}_{\mathcal{B}}$ is a proper subcategory of $\mathcal{K}$. There are other class of objects in $\mathrm{ob}(\mathcal{K})$, which are obtained by a \textit{bi-morphism} $\mathfrak{b}_{\mathcal{K} \mathcal{B}} : \mathrm{ob}(\mathcal{K}_{\mathcal{B}}) \mapsto \mathrm{ob}(\mathcal{K}_{\mathrm{E}})$ such that 
\be 
\exists \, \mathfrak{b}_{\mathcal{K}_{\mathrm{E}} \mathcal{B}}, \,\, \mathfrak{b}_{\mathcal{K}_{\mathrm{E}} \mathcal{B}}: X_{\mathcal{K}_{\mathcal{B}}} \mapsto Y_{\mathcal{K}_{\mathrm{E}}} \quad \mathrm{iff} \quad X=Y\,, \quad \forall \, X \in \mathrm{ob}(\mathcal{K}_{\mathcal{B}})  \quad \mathrm{and} \quad \forall\, Y \in \mathrm{ob}(\mathcal{K}_{\mathrm{E}})\,,
\ee
where the subcategory $\mathcal{K}_{\mathrm{E}}$ is the complement category of $\mathcal{K}_{\mathrm{B}}$ and the objects of $\mathrm{ob}(\mathcal{K}_{\mathrm{E}})$ are denoted as
\be
\mathrm{ob}(\mathcal{K}_{\mathrm{E}}): \,\, \mathcal{R}_{\mathcal{K}_{\mathrm{E}}}, \mathcal{F}_{\mathcal{K}_{\mathrm{E}}}, \mathcal{(NS)}_{\mathcal{K}_{\mathrm{E}}} \,.
\ee

The morphism $\mathfrak{b}_{\mathcal{K}_{\mathrm{E}} \mathcal{B}}$ is an operation, which engenders an ontological mode of beable $\mathcal{B}$ as an epistemological mode of knowable $\mathcal{K}$. Specifically, $\mathfrak{b}_{\mathcal{K}_{\mathrm{E}} \mathcal{B}}$ functions by forgetting the object $X_{\mathcal{K}_{\mathcal{B}}}$ as in $\mathrm{ob}(\mathcal{K}_{\mathcal{B}})$ and reassigning it as an object in $\mathrm{ob}(\mathcal{K}_{\mathrm{E}})$, while retaining its universal onto-epistemological mode $X$. Consequently, each mode of $\mathcal{B}$ has a distinct corresponding mode in $\mathcal{K}$. It should be noted that $\mathfrak{b}_{\mathcal{K}_{\mathrm{E}} \mathcal{B}}$ is not an isomorphism, as an inverse of $\mathfrak{b}_{\mathcal{K}_{\mathrm{E}} \mathcal{B}}$ is not required. This is due to the postulated non-equivalence between existential modes and modes of knowledge.

There exists another class of morphisms in the subcategory $\mathcal{K}_{\mathrm{E}}$: $\mathfrak{f}_{X_{\mathcal{K}_{\mathrm{E}}}, Y_{\mathcal{K}_{\mathrm{E}}}}:  X_{\mathcal{K}_{\mathrm{E}}} \mapsto Y_{\mathcal{K}_{\mathrm{E}}}$ and $\mathfrak{g}_{X_{\mathcal{K}_{\mathrm{E}}}, Y_{\mathcal{K}_{\mathrm{E}}}}:  Y_{\mathcal{K}_{\mathrm{E}}} \mapsto X_{\mathcal{K}_{\mathrm{E}}}\,, \, \forall\, X_{\mathcal{K}_{\mathrm{E}}}, Y_{\mathcal{K}_{\mathrm{E}}} \in \mathrm{ob}(\mathcal{K}_{\mathrm{E}})$ such that  
\be
\mathfrak{f}_{X_{\mathcal{K}_{\mathrm{E}}}, Y_{\mathcal{K}_{\mathrm{E}}}} \circ \mathfrak{g}_{X_{\mathcal{K}_{\mathrm{E}}}, Y_{\mathcal{K}_{\mathrm{E}}}} = \mathrm{id}_{X_{\mathcal{K}_{\mathrm{E}}}} \quad \mathrm{and} \quad \mathfrak{g}_{X_{\mathcal{K}_{\mathrm{E}}}, Y_{\mathcal{K}_{\mathrm{E}}}} \circ \mathfrak{f}_{X_{\mathcal{K}_{\mathrm{E}}}, Y_{\mathcal{K}_{\mathrm{E}}}} = \mathrm{id}_{Y_{\mathcal{K}_{\mathrm{E}}}}\,, 
\ee
where the isomorphisms $\mathfrak{f}_{X_{\mathcal{K}_{\mathrm{E}}}, Y_{\mathcal{K}_{\mathrm{E}}}}$ and $\mathfrak{g}_{X_{\mathcal{K}_{\mathrm{E}}}, Y_{\mathcal{K}_{\mathrm{E}}}}$ indicate that any mode of $\mathcal{K}_{\mathrm{E}}$ implies the existence of all other modes of $\mathcal{K}_{\mathrm{E}}$: $\mathcal{R}_{\mathcal{K}_{\mathrm{E}}} \cong \mathcal{F}_{\mathcal{K}_{\mathrm{E}}} \cong \mathcal{(NS)}_{\mathcal{K}_{\mathrm{E}}}$.

The categories $\mathrm{cat}(\mathcal{NK})$ of non-knowable ($\mathcal{NK}$) arise from the action of the abstraction functor $\mathfrak{a}$ on the objects and morphisms in $\mathcal{K}_{\mathrm{E}}$, analogous to the formation of non-beable $\mathcal{NB}$. Indeed, there exists a full and faithful functor $F_{\mathcal{NB}, \, \mathcal{NK}}$ that induces an isomorphism between the categories $\mathcal{X}_{\mathcal{NB}} \in \mathrm{cat}(\mathcal{NB})$ and $\mathcal{X}_{\mathcal{NK}} \in \mathrm{cat}(\mathcal{NK})$: $\mathcal{X}_{\mathcal{NB}} \cong \mathcal{X}_{\mathcal{NK}}$, including $\varnothing_{\mathcal{NK}} \equiv \varnothing_{\mathcal{NB}}$. Categorically, there is no distinction between references to the non-knowable and non-beable.

\subsection{Universal set of onto-epistemology of $\mathcal{RM}$}

Now, let us define a master set (or master topology) $S_{\mathcal{RM}}$ as the power set $\mathfrak{p}(\bigcup_{\alpha} S_{\alpha})$ of the union of the sets of all objects in $\mathcal{B}, \mathcal{K}_{\mathrm{E}}$ and $\mathcal{NB}$:
\be
S_{\mathcal{RM}} := \mathfrak{p} \, \big(\bigcup_{\alpha} S_{\alpha}\big) = \{ \, \{\cdots, x_i ,\cdots \} \,\, | \,\, \forall x_i \in S_{\mathcal{B}} \,\, \mathrm{or} \,\, \forall x_i  \in S_{\mathcal{K}_{\mathrm{E}}} \,\, \mathrm{or} \,\,  \forall x_i  \in S_{\mathcal{NB}} \} \,,
\ee
where the sets $S_{\mathcal{B}}, S_{\mathcal{K}_{\mathrm{E}}}$ and $S_{\mathcal{NB}}$ are:
\begin{align}
S_{\mathcal{B}} = \{ \mathcal{R}_{\mathcal{B}}, \mathcal{F}_{\mathcal{B}}, (\mathcal{NS})_{\mathcal{B}} \}, \quad S_{\mathcal{K}_{\mathrm{E}}} = \{ \mathcal{R}_{\mathcal{K}_{\mathrm{E}}}, \mathcal{F}_{\mathcal{K}_{\mathrm{E}}}, (\mathcal{NS})_{\mathcal{K}_{\mathrm{E}}} \}\, \quad S_{\mathcal{NB}} = \{ \mathcal{A}, \mathcal{F}, \mathcal{NS} \} \,.
\end{align}

The universal set $U_{\mathcal{RM}}$ of the onto-epistemology of $\mathcal{RM}$, which constitutes the existential domain of $\mathcal{RM}$ and represents the coexistence of the universal modes, is realized as the non-empty subset $U_{\mathcal{RM}} \subset S_{\mathcal{B}} \times S_{\mathcal{K}_{\mathrm{E}}} \subset S_{\mathcal{RM}}$ satisfying the uniformity of each pair of universal modes of existence and knowledge as its element:
\be
U_{\mathcal{RM}} := \bigcup_i \, S_{X_i}\,, \quad S_{X_i} = \{ X_{\mathcal{B}}, X_{\mathcal{K}_{\mathrm{E}}} \}_i \,, \quad \forall X_{\mathcal{B}} \in S_{\mathcal{B}} \quad \mathrm{and} \quad \forall X_{\mathcal{K}_{\mathrm{E}}} \in S_{\mathcal{K}_{\mathrm{E}}}\,,
\ee
which is indeed covered by the subsets $S_X \subset U_{\mathcal{RM}}$ associated with each of the universal onto-epistemological modes $X: \mathcal{R}, \, \mathcal{F}, \, \mathcal{NS}$. The fact that the elements of the disjoint subsets $S_X$ belong to the same kind is due to the ontological irreducibility of universal onto-epistemological modes that there exists no functors between the universal subsets $\{S_X\}$ associated to each universal mode, except the presence of morphism between the elements of each $S_X$. Conversely, the complement of $U_{\mathcal{RM}}$ refers to the irrelevant and non-existential domain of $\mathcal{RM}$. Overall, the universal set $U_{\mathcal{RM}}$ will function as \textit{organizing center} within the space of sets of onto-epistemological particularities in $\mathcal{RM}$.

\subsection{Onto-epistemological particularities in $\mathcal{RM}$}

Let $\mathcal{P}_j$ be a category in the indexed family $\{\mathcal{P}_{j}\}_{j \in J}$ of onto-epistemological particularities $\mathcal{P}$, where $J \subset \mathbb{Z}^+$ is a bounded (not closed) index set providing a specific enumeration of $\mathcal{P}$. The ontological ($\mathcal{B}$) and epistemological ($\mathcal{K}$) contents of $\mathcal{P}_{j}$, denoted by $\mathcal{P}_{\mathcal{B}, j}$ and $\mathcal{P}_{\mathcal{K}, j}$, are designated by a secondary index set $\mathcal{A} \in \{\mathcal{B}, \mathcal{K}\}$.

The ontological subcategory $\mathcal{P}_{\mathcal{B}, j}$ is defined via an injection functor $I_{\mathcal{B} \mathcal{P}_j}$ that maps the objects of $\mathcal{B}$ into the objects of the category $\mathcal{P}_{\mathcal{B}, j}$, such that $I_{\mathcal{B} \mathcal{P}_j}$ maps $\mathcal{B}$ to its equivalent full subcategory $I_{\mathcal{P}_j} \mathcal{B}$ in $\mathcal{P}_{\mathcal{B}, j}$:
\be
\exists \, I_{\mathcal{B} \mathcal{P}_j}, \,\,\,\,  I_{\mathcal{B} \mathcal{P}_j}\!: \,\, \mathrm{ob}(\mathcal{B}) \mapsto \mathrm{ob}(I_{\mathcal{P}_j} \mathcal{B})
\ee 
such that for all $X \in \mathrm{ob}(\mathcal{B})$, we have the following morphisms in the subcategory $I_{\mathcal{P}_j} \mathcal{B}$:
\be
m_{\mathcal{P}_{\mathcal{B}, j}}(I_{\mathcal{B} \mathcal{P}_j}(X), I_{\mathcal{B} \mathcal{P}_j}(X)) = \mathrm{id}_{I_{\mathcal{B} \mathcal{P}_j}(X)}\,.
\ee
Therefore, the particularity $\mathcal{P}_{\mathcal{B}, j}$ can be considered a \textit{particular beable}, as the universal modes of $\mathcal{B}$ are embedded into $\mathcal{P}_{\mathcal{B}, j}$: 
\be
\mathrm{ob}(I_{\mathcal{P}_j} \mathcal{B}): \,\,\, \mathcal{R}_{\mathcal{P}_{\mathcal{B},j}} , \,\,  \mathcal{F}_{\mathcal{P}_{\mathcal{B},j}}, \,\, \mathcal{(NS)}_{\mathcal{P}_{\mathcal{B},j}} \,.
\ee

The subcategory $I_{\mathcal{P}_j} \mathcal{B}$ is a proper subcategory of $\mathcal{P}_{\mathcal{B}, j}$. There exists another class of objects in $\mathrm{ob}(\mathcal{P}_{\mathcal{B}, j})$, which are obtained by a \textit{attaching epi-morphism} $\mathfrak{a}_{\mathcal{P}_{\mathcal{B}, j}}$ such that 
\be 
\mathfrak{a}_{\mathcal{P}_{\mathcal{B}, j}} \! : X_{\mathcal{P}_{\mathcal{B},j}}  \mapsto Y_{\{X_{\mathcal{P}_{\mathcal{B},j}} \}} \,, \,\,\, \forall \, X_{\mathcal{P}_{\mathcal{B},j}} \in \mathrm{ob}(I_{\mathcal{P}_j} \mathcal{B})  \,\,\, \mathrm{and} \,\,\, Y_{\{X_{\mathcal{P}_{\mathcal{B},j}} \}} \in \mathrm{ob}(\mathcal{P}^{(\mathfrak{a})}_{\mathcal{B}, j}) \,,
\ee
i.e., each object in $\mathrm{ob}(I_{\mathcal{P}_j} \mathcal{B}) $ is mapped to every object in $\mathrm{ob}(\mathcal{P}_{\mathcal{B}, j})$ that is not an object of $I_{\mathcal{P}_j} \mathcal{B}$. The objects obtained by $\mathfrak{a}_{\mathcal{P}_{\mathcal{B}, j}}$ form a discrete subcategory, denoted by $\mathcal{P}^{(\mathfrak{a})}_{\mathcal{B}, j}$, where the only morphisms between its objects are identity maps. The morphism $\mathfrak{a}_{\mathcal{P}_{\mathcal{B}, j}}$ operates by assigning a \textit{multiplicity} to each element $Y_{\{X_{\mathcal{P}_{\mathcal{B},j}} \}} \in \mathrm{ob}(\mathcal{P}^{(\mathfrak{a})}_{\mathcal{B}, j})$. In other words, every $Y_{\{X_{\mathcal{P}_{\mathcal{B},j}} \}}$ of $\mathcal{P}^{(\mathfrak{a})}_{\mathcal{B}, j}$ is essentially attached with the (isomorphically equivalent) universal modes $\{X_{\mathcal{B}}\}$ of $\mathcal{B}$.

The ontological content of the particularity $\mathcal{P}_{j}$ corresponds to the epistemological content $\mathcal{P}_{\mathcal{K}, j}$ through the injection functor $I_{\mathcal{P}_{\mathcal{B}, j} \mathcal{P}_{\mathcal{K}, j}}$. This functor embeds all objects $Y_{\{X_{\mathcal{P}_{\mathcal{B},j}} \}} \in \mathrm{ob}(\mathcal{P}^{(\mathfrak{a})}_{\mathcal{B}, j})$ into a full subcategory $\mathcal{P}^{(\mathfrak{a})}_{\mathcal{K}_{\mathcal{B}}, j}$ of $\mathcal{P}_{\mathcal{K}, j}$. The objects in $\mathcal{P}^{(\mathfrak{a})}_{\mathcal{K}_{\mathcal{B}}, j}$ of $\mathcal{P}_{\mathcal{K}, j}$, which are isomorphically equivalent to the objects $\mathrm{ob}(\mathcal{P}^{(\mathfrak{a})}_{\mathcal{B}, j})$, are then mapped by a bimorphism $\mathfrak{b}_{\mathcal{P}_{\mathcal{K}_{\mathrm{E}}} \mathcal{P}_{\mathcal{K}_{\mathrm{B}}}}: \mathcal{P}^{(\mathfrak{a})}_{\mathcal{K}_{\mathcal{B}}, j} \mapsto  \mathcal{P}^{(\mathfrak{a})}_{\mathcal{K}_{\mathrm{E}}, j}$. Similar to $\mathfrak{b}_{\mathcal{K}_{\mathrm{E}} \mathcal{B}}$, the morphism $\mathfrak{b}_{\mathcal{P}_{\mathcal{K}_{\mathrm{E}}} \mathcal{P}_{\mathcal{K}_{\mathrm{B}}}}$ reassociates any object $X$ as in $\mathrm{ob}(\mathcal{P}^{(\mathfrak{a})}_{\mathcal{K}_{\mathcal{B}}, j})$ as an object in $\mathrm{ob}(\mathcal{P}^{(\mathfrak{a})}_{\mathcal{K}_{\mathrm{E}}, j})$, while preserving its association wtih the onto-epistemological particularity $\mathcal{P}_j$.

The formation of the objects in the subcategory $\mathcal{P}^{(\mathfrak{a})}_{\mathcal{K}_{\mathrm{E}}, j}$ is equally well-defined by introducing an injection functor $I_{\mathcal{K}_{\mathrm{E}} \mathcal{P}_j}$, which embeds the subcategory $\mathcal{K}_{\mathrm{E}}$ as a full subcategory $I_{\mathcal{P}_j} \mathcal{K}_{\mathrm{E}}$ in $\mathcal{P}_{\mathcal{K}, j}$. This is followed by the application of the attaching epimorphism $\mathfrak{a}_{\mathcal{P}_{\mathcal{K}_{\mathrm{E}}, j}}$ to finally establish the subcategory $\mathcal{P}^{(\mathfrak{a})}_{\mathcal{K}_{\mathrm{E}}, j}$.

The coexistence of onto-epistemological particularities is formulated through a set $S_{\mathcal{P}_j}$ for each onto-epistemological particularity $\mathcal{P}_j$:
\be
S_{\mathcal{P}_j} := \big(\bigcup_{\beta} \, S_{Y_{\beta}} \big) \, \bigcup \, U_{\mathcal{RM}}\,, \quad  S_{Y_{\beta}}  \equiv \{ Y_{\{X_{\mathcal{P}_{\mathcal{B},j}} \}} , Y_{\{X_{\mathcal{P}_{\mathcal{K}_{\mathrm{E}},j}} \}} \}_{\beta}
\ee
such that the universal set $U_{\mathcal{RM}}$, a subset of $S_{\mathcal{P}_j}$, is to provide universal modes of existence and knowledge within $S_{\mathcal{P}_j}$.

Consequently, the space of onto-epistemological sets in $\mathcal{RM}$ is to be characterized by the union of all particular sets $S_{\mathcal{P}_j}$ and their common subset $U_{\mathcal{RM}}$:
\be
S_{\mathcal{P}_{\mathcal{RM}}} = \bigcup_j \, S_{\mathcal{P}_j}\,, \quad  U_{\mathcal{RM}} = \bigcap_j S_{\mathcal{P}_j} \,.
\ee

Let us list the onto-epistemological particularities with assigned (ontological/epistemological) $( \cdot / \cdot)$ units and some of their associated objects:

\begin{itemize}
 \item[i.] \textbf{Interactibility} (\textit{Effect/Phenomenon}): \textit{Locality, Agency, Manifestibility, etc.}
 \item[ii.] \textbf{Transformability}  (\textit{Process/Transition}):  \textit{Reversibility/Irreversibility, Symmetry/Asymmetry, etc.}
\item[iii.] \textbf{Structurability}  (\textit{Structure/System}): \textit{Regularity, Articularity, Stratification, etc.} 
\item[iv.] \textbf{Scale-dependency} (\textit{Base/Measure}): \textit{Irreducibility, Boundaries, Measurability, etc.}
\item[v.] \textbf{Contextuality} (\textit{Information/Perspective}): \textit{Indexicality, Commensurability, Fragmentability, etc.}
\item[vi.]  \textbf{Actuality}: (\textit{Act/State}): \textit{Eventuality, Potentiality, Observability, etc.}
\item[vii.]\textbf{Contingency}: (\textit{Relata/Probability}): \textit{Indeterminancy, Causality, Predictibility, etc.}
\end{itemize}

\begin{figure}[htbp]
\centering
 \includegraphics[scale=0.8]{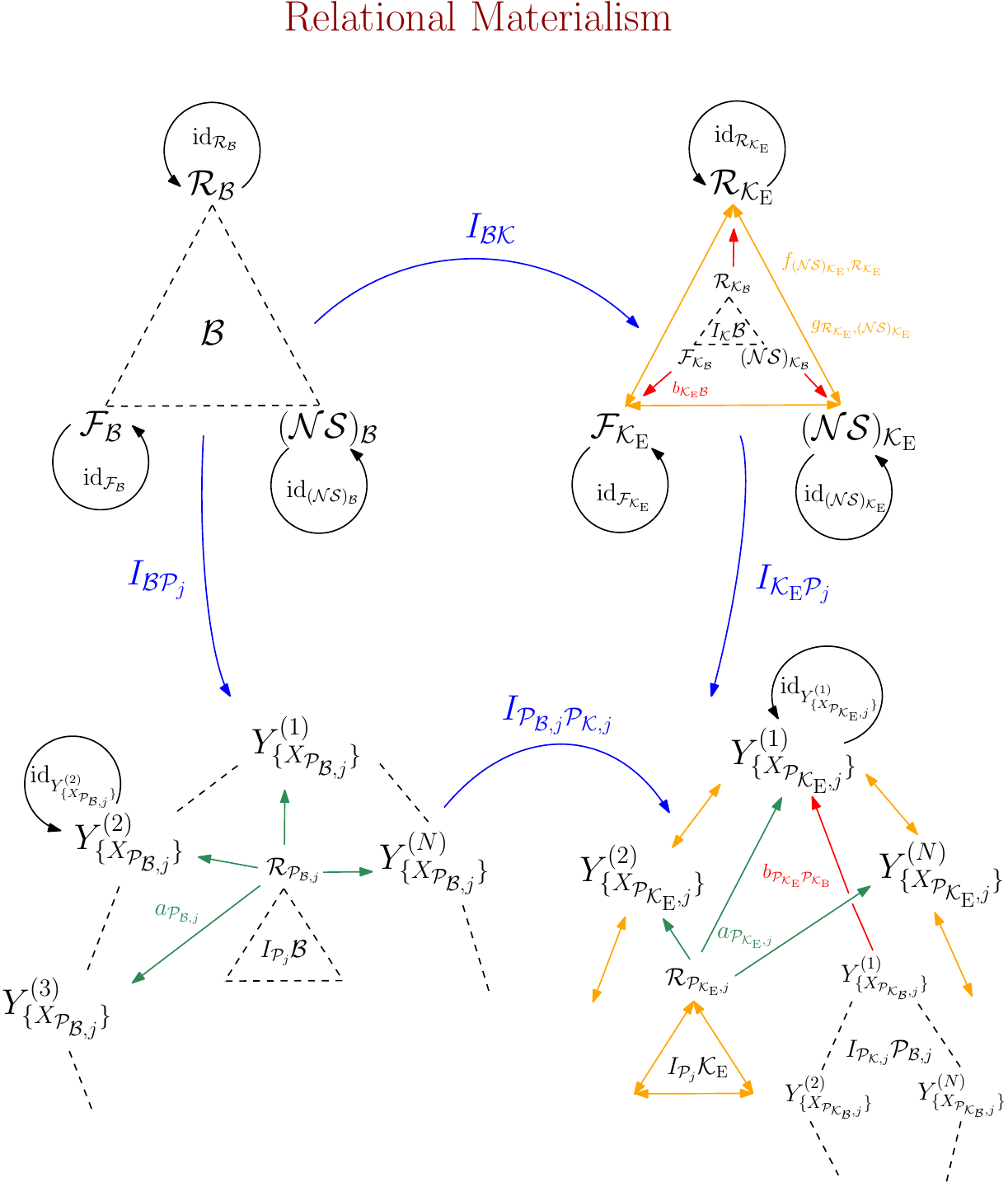}
\caption{Onto-epistemology of $\mathcal{RM}$.}
\label{fig:diagrammRM}
\end{figure}

A diagrammatic representation of relational materialism is presented in Fig.~\ref{fig:diagrammRM}.

\section{Conclusion}
\label{sec:concl}

Relational materialism is a philosophical perspective aimed at establishing primary principles and concepts pertinent to both existence and knowledge. The modes of existence and knowledge function as characterizations of the probable ways entities can exist or be known. These modes are categorized into universal and particular ontological modes, each with a corresponding epistemological mode. The materiality of these onto-epistemological modes is contingent upon their coexistential nature. The identification of these modes is not a priori; rather, they are derived from our experiences and practices, which are consolidated throughout the scientific journey.

We propose an open and dynamical framework wherein neither universals nor particularities are fixed permanently. Instead, the list of modes can be extended, universal and particular modes can be interchanged, or some modes may be eliminated, depending on our evolving capacity to comprehend the reality around us. Below is a list of references, which, though likely incomplete, have served as sources of inspiration and enrichment.

\end{document}